\documentstyle[12pt]{article}

\begin{document}

\baselineskip=7mm

\begin{large}

\begin{center}

{Ground state spin 0$^+$ dominance of many-body systems with 
random interactions and related topics }

\vspace{0.1in}
    
 A. Arima$^1$, N. Yoshinaga$^2$, and    Y. M.  Zhao$^{2,3}$

\vspace{0.2in}
$1$)The House of Councilors, 2-1-1 Nagatacho,

   Chiyodaku, Tokyo 100-8962,  Japan

$2$) Department of Physics,
     Saitama University, 338 Saitama Japan

$3$) Department of Physics, Southeast University,
        Nanjing 210018, P. R. China.

\vspace{0.315in}

\end{center}
\end{large}

\vspace{0.1in}

\begin{small}

In this talk we shall show our recent results in understanding
the spin$^{\rm parity}$ 0$^+$ ground state (0 g.s.) 
dominance of many-body systems. We propose a simple 
approach to predict the spin $I$ g.s. probabilities which does not 
require the diagonalization of a Hamiltonian with random interactions. Some 
findings related to the 0 g.s. dominance will also be discussed.

\end{small}

\newpage

\section{INTRODUCTION}

The ground state spins$^{\rm parity}$ of even-even nuclei are always 0$^+$, 
which is believed to be a consequence of the attractive short-range 
interactions between nucleons. However, 
a predominance of spin$^{\rm parity}$ 0$^+$ ground states (0 g.s.)
 was discovered by Johnson, Bertsch and Dean in 1998 using 
the two-body random ensemble (TBRE) \cite{Johnson1} and was related to
a reminiscence of generalized seniority by Johnson, Bertsch,
Dean and Talmi in 1999 \cite{Johnson2}. These phenomena have been
confirmed by many works in different systems \cite{Bijker0}. There
have been many  efforts to understand
this interesting and important observation
\cite{Bijker1,Bijker2,Kaplan,Zhao2,Zhao3,Zhao3-1}. Further works have been
reported on generic collectivity of many-body systems in the presence 
of random interactions \cite{Bijker1,Zhao4},
the behavior of average energies \cite{Zhao3,Zhao3-1}, etc.

\section{SIMPLE SYSTEMS WITH A FEW  FERMIONS IN SINGLE-$j$ SHELLS}

\subsection{Results obtained by diagonalizing energy matrices with the TBRE}

The Hamiltonian that we use for fermions in a single-$j$ shell is defined as follows
\begin{eqnarray}
&& H = \sum_J G_J  A^{J \dagger} \cdot A^{J} \equiv  
\sum_J \sqrt{2J+1} G_J 
\left( A^{J \dagger} \times
\tilde{A}^J \right)^0, \nonumber \\
&&  A^{J \dagger} = \frac{1}{\sqrt{2}} \left( a_j^{\dagger} \times a_j^{\dagger}
 \right)^J, ~ ~
 \tilde{A}^J = - \frac{1}{\sqrt{2}} \left( \tilde{a}_j \times
 \tilde{a}_j \right)^J, ~ ~ G_J = \langle j^2 J|V| j^2 J \rangle.  \label{pair}
  \nonumber
\end{eqnarray}
$G_J$'s are taken as a
set of Gaussian-type random numbers with  
a width being 1 and an average being 0. This two-body random ensemble
is referred to as ``TBRE". 
 Hamiltonian of fermions in many-$j$ shells
or a  boson Hamiltonian can be defined in a similar way.
The $I$ g.s. probabilities in this paper are obtained by 1000 runs
of a TBRE Hamiltonian.

We take very simple systems such as those with four fermions  in a
single-$j$ shell. 
Many different values of $j$ from 7/2 to 33/2 are taken
into account. Fig. 1 shows the 0 g.s. probabilities of 4 fermions
in different single-$j$ shells. 
 From this figure we observe interesting
oscillations of the 0 g.s. probabilities $P(0)$ as a function of $j$.  
One also notices easily that  the $P(0)$'s are the largest, 
except for a few small $j$ cases among
the $P(I)$'s, which are the probabilities of a state with spin $I$ to be the 
ground state.

\subsection{An empirical formula to predict the $P(I)$'s}

Let us set only one of the $G_J$'s equal to $-1$ and the others to zero, 
and find the spin $I$ of the ground state. We 
repeat this process for 
all two-body interactions $G_J$. 
We can find how many times the ground state has angular momentum $I$. 
This number is denoted as $N_I$ and the values of $N_I$'s
for four nucleons in a single-$j$ shell can be
easily counted by looking at Table 1.
Using the $N_I$, we can predict the probability that the ground state 
has angular momentum $I$ as $P(I)_{emp} = N_I/N$, where $N$ is the number of 
independent two-body matrix elements $(N=j+1/2$ for fermions in a 
single-$j$ shell).

A nice agreement between our predicted $P(0)$'s and
those obtained by diagonalizing a TBRE Hamiltonian is shown
in Fig. 2. 
The empirical formula is also used to predict 
$P(I)$'s of systems with four and 
five nucleons in a $j=9/2$ shell. The results are shown in Fig. 3 a) and b). 
The agreements are again remarkable. 
For more complicated systems, such as those with four,
five, six and seven nucleons in two-$j$
shells and $sd$ bosons, the formula works very well, too. To examplify, 
Fig. 3c) and 3d) show the cases of  seven fermions in a
two-$j$ ($j_1=7/2$, $j_2=5/2$) shell and 10 $sd$ bosons.

\section{AVERAGE ENERGIES}

\subsection{Probabilities of $I$ g.s. for average energies $\bar{E}_I$}

Our problem becomes simpler for the  complicated systems discussed 
in the previous sections, if we take a trace of each energy matrix. 
The average energy $\bar{E}_I$ can be expressed in terms of
linear combination of $G_J$'s:
\begin{equation}
\bar{E}_I = \sum_J \bar{\alpha}^J_I G_J,
\end{equation}
where $\bar{\alpha}^J_I$ is obtained by averaging 
\begin{eqnarray}
\alpha_{\beta I}^J= \frac{n(n-1) }{2} \sum_{K, \gamma} \left(
\langle j^{n-2} K \gamma, j^2 J| \}
j^n I \beta \rangle \right)^2 \nonumber 
\end{eqnarray}
over all $\beta$'s. Here $\langle j^{n-2} K \gamma, j^2 J| \} j^n I \beta \rangle $ are the
two-body coefficients of fractional parentage, and $\beta$  
(or $\gamma$) refers to
additional quantum numbers to define a state of $n$ (or $n-2$)
fermions with total angular momentum $I$ (or $K)$ uniquely.

We are now ready to apply the same method to predict the probabilities 
${\cal P}(I)$'s of  $\bar{E}_I$ to be  the lowest.  From Table 2,
we can find that 
$N_0=1$, $N_2 =1$, $N_3=2$ and $N_{12}=1$ for
4 fermions in a $j=\frac{9}{2}$ shell. The total number $N$ of
$G_J$'s in this shell is 5. Using the empirical formula  
discussed above, we  can predict the ${\cal P}(I)$'s. The results 
(labeled by ``${\cal P}^{G_J=-1}(I)$") are 
shown in  Table 3 together with ${\cal P}^{exp}(I)$'s 
calculated by diagonalizing
a TBRE Hamiltonian in 1000 runs.

Comparing   the values of ${\cal P}^{exp}(I)$'s  with the ones shown as 
${\cal P}^{G_J=-1}(I)$, we see that the agreements are qualitatively good.
However, they are not as
good as in the cases discussed in  the previous section. 

One thing, here, should be examined. When the sign of $G_J$ is changed to 
be positive, again some $\bar{E}_I$'s can be the lowest. Because all 
$\bar{E}_I$'s are pushed up, the lowest ones are those which are least 
favored by $G_{J'} = \delta_{J'J}$. Table 2 shows the lowest $I$'s
which are  given by $G_{J'} = \delta_{J'J}$ (the row labeled
by ``$G_J = + 1$").

For $G_0$ there are several $\bar{E}_I$'s with $I=$3, 5, 7, 9, 10, 12 are 
degenerate. When this kind of degeneracy occurs, we give each of them 
a weight (one over how many times are degenerate). We then count how many 
$G_{J'} = \delta_{J'J}$ and $G_{J'} = -\delta_{J'J}$ give the lowest 
energy to an angular momentum $I$. We divide this number by $2j+1$ which 
is the total number of $G_{J'} = -\delta_{J'J}$ and $G_{J'} = 
\delta_{J'J}$. The probabilities thus calculated are shown as 
${\cal P}^{G_J = \pm 1}$. We see much better agreements between the 
${\cal P}^{exp}$ calculated  using a TBRE and
those shown as ${\cal P}^{G_J = \pm 1}$.

We then have to reconsider the empirical way to predict $P(I)$'s, which 
is discussed in the previous section, where we take into account $G_{J'} = 
-\delta_{J'J}$ only. The same kind of improvement can be 
obtained, as shown  in Table 5. However, a large number of states have roughly the same energy near the ground 
states, and therefore the results are not so different
from the prediction when
only $G_{J'} = -\delta_{J'J}$ is taken into account.

\subsection{A trajectory of average energy in terms of $I(I+1)$}

Let $\langle \bar{E}_I \rangle_{min}$
be a quantity obtained by averaging $\bar{E}_I$
only over those cases where $I$ of the lowest $\bar{E}_I$ is 
 around $I_{min}$ among the
ensemble. We find that $\langle \bar{E}_I \rangle_{min}$
is nearly proportional to $I(I+1)$,
similar to a rotational spectra. The very same can be found when we replace ``min" 
by ``max". This is because ${\cal P}(I)$'s are roughly symmetric about 
$I_{max}/2$.
Several examples of
$\langle \bar{E}_I \rangle_{min}$ are shown in Fig. 4.

We can explain this finding statistically. The details are described in 
\cite{Zhao3-1}.

\section{BY-PRODUCTS}

As having been well known for many years, the monopole pairing
$G_J= -\delta_{J0}$ gives a ground state with $I=0$ for an even
number of fermions and a ground state with $I=j$ for an odd number of fermions.
It is quite interesting to observe that $G_J= -\delta_{J2}$
gives a ground state with $I=n$ ($n$=even) and
$I=j-(n-1)/2$ ($n=$odd) in most cases that we have checked \cite{Zhao2}, 
although  there are some exceptions.
Here $n$ is the number of fermions in a single-$j$ shell.

We also observe that the interaction of $G=-\delta_{J (2j-1)}$
gives a large array of eigenvalues which are
asymptotic integers. An understanding of this observation is
in progress \cite{Zhao_joe}.

\section{SUMMARY}

A simple method is proposed to predict the probability $P(I)$ that the
ground state has angular momentum $I$ in many-body systems interacting
via a two-body random ensemble. We find and predict that $P(0)$ is always
the largest except for a few cases.

We also study the probabilities of average energies with fixed $I$ to be
the lowest.  It is interesting to find a trajectory of the energy
$\langle \bar{E}_I \rangle_{min}$
(and $\langle \bar{E}_I \rangle_{max}$) which is nearly proportional to
$I(I+1)$.

As by-products, we find: 1) the quadrupole pairing interaction
seems to favor the ground states which have angular momentum $n$ ($n$ is even)
or $j-(n-1)/2$ ($n$ is odd) in systems with $n$ fermions
in a single-$j$ shell; 2) the highest-multipole pairing interaction
presents a large array of asymptotic integer eigenvalues.

\newpage

Figure Captions:

\vspace{0.3in}

Figure ~ 1 ~~ 
The g.s.  probabilities of $I=$0, 2, 4, and $I_{max}$
of four fermions in a single-$j$ shell.

\vspace{0.3in}

Figure ~ 2 ~~ 
A comparison between our predicted $P(0)$'s and those
obtained by 1000 runs of a TBRE Hamiltonian.

\vspace{0.3in}

Figure ~ 3 ~~
Several examples of comparison between the empirically predicted 
  $I$ g.s.  probabilities and those obtained by diagonalizing TBRE Hamiltonians.

\vspace{0.3in}

Figure ~ 4 ~~
The $I(I+1)$ behavior of $\langle \bar{E} \rangle_{min}$.

\newpage

Table ~ 1 ~~
{  The angular momenta $I$'s which give the lowest eigenvalues
for 4 fermions in a single-$j$ shell, 
when  $G_J=-1$  and all other parameters are  0.    }

\vspace{0.2in}

\begin{tiny}
\begin{tabular}{ccccccccccccccccc} \hline  
$2j$ &  $G_0$ &  $G_2$ &  $G_4$ &  $G_6$ &  $G_8$ &  $G_{10}$ &  $G_{12}$
&  $G_{14}$ &  $G_{16}$ &  $G_{18}$ &  $G_{20}$ &  $G_{22}$ &  $G_{24}$
&  $G_{26}$ &  $G_{28}$ &  $G_{30}$   \\  \hline
7  & 0 &4 &2 &8 &   &   & & & &  & & & & & & \\
9  & 0 &4 &0 &0 &12 &   & & & &  & & & & & & \\
11 & 0 &4 &0 &4 &8  &16 & & & &  & & & & & & \\
13 & 0 &4 &0 &2 &2  &12 &20 & & &  & & & & & & \\
15 & 0 &4 &0 &2 &0  &0  &16 &24 & &  & & & & & & \\
17 & 0 &4 &6 &0 &4  &2  &0  &20 &28 &  & & & & & & \\
19 & 0 &4 &8 &0 &2  &8  &2  &16 &24 &32 & & & & & & \\
21 & 0 &4 &8 &0 &2  &0  &0  &0  &20 &28 &36 & & & & & \\
23 & 0 &4 &8 &0 &2  &0  &10 &2  &0  &24 &32 &40 & & & & \\
25 & 0 &4 &8 &0 &2  &4  &8  &10 &6  &0  &28 &36 &44 & & & \\
27 & 0 &4 &8 &0 &2  &4  &2  &0  &0  &4  &20 &32 &40 &48 & & \\
29 & 0 &4 &8 &0 &0  &2  &6  &8  &12 &8  &0  &24 &36 &44 &52 & \\
31 & 0 &4 &8 &0 &0  &2  &0  &8  &14 &16 &6  &0  &32 &40 &48 & 56 \\
  \hline 
\end{tabular}
\end{tiny}

\vspace{0.4in}

Table ~ 2 ~~ 
{ The  spins of the lowest $\bar{E}_I$ for 4 fermions in a single 
$j$=9/2 shell when only one of $G_J$ is set to $-1$ or $+1$.}

\vspace{0.2in}

\begin{small}
 \begin{tabular}{cccccc}
\hline
            &  $G_0$  &  $G_2$  &  $G_4$  &  $G_6$   &
$G_8$  \\   \hline
$G_J=-1$    &  0 &  2 &   3 &  3  &  12 \\
$G_J=+1$    & ~~~~~~~ 3, 5, 7, 9, 10,12 ~~~~~~
& ~~~~    12 ~~~~ & ~~  12 ~~ & ~~ 2 ~~ & ~~ 3 ~~ \\ 
\hline
\end{tabular}\\[2pt]
\end{small}  

\vspace{0.4in}

Table ~ 3 ~~
{{\em I} g.s. probabilities for average energies of four
fermions in a {\em j}=9/2 shell.}

\vspace{0.2in}

\begin{tiny}
\begin{tabular}{llllllllllll}
\hline
$I$           &  $0~~~$  &  $2~~~$  &  $3~~~$  &  $4~~~$   &
$5~~~$           & $6~~~$ & $7~~~$  &  $8~~~$  &  $9~~~$   & 
$10~~$          & $12~~$ \\
\hline
${\cal P}^{exp}(I)$          &  10.2 &  15.4 &  28.9 &  1.7~~ &
0.6~                       &  0.3~~  &  3.2~  &  0~~~    &  0~~~  &
8.7~                       &  31.0 \\
${\cal P}^{G_J=-1}(I)$          &  20 &  20 &  40 &  0~~~ &
0~~~~                       &  0~~~~  &  0~~~~  &  0~~~    &  0~~~  &
0~~~~                       &  20~~ \\
${\cal P}^{G_J=\pm 1}(I)$          &  10~~ &  21.6 &  31.6 & 0~~~ &
1.6~~                       &  0~~~  &  1.6~  &  0~~~    &  1.6~~  &
1.6~                       &  31.6 \\  
\hline
\end{tabular}\\[2pt]
\end{tiny}

\newpage

\vspace{0.4in}

Table ~ 4 ~~ 
{ The ground state spins of 4 fermions in a single 
{\em j}=9/2 shell when only one of $G_J$ is set to $+1$ or $-1$.}

\vspace{0.2in}

\begin{tiny}
\begin{tabular}{cccccc}
\hline
            &  $G_0$  &  $G_2$  &  $G_4$  &  $G_6$   &
$G_8$  \\   \hline
$G_J=-1$    &  0 &  4 &   0 &  0  &  12 \\
$G_J=+1$    & ~~~~~~~ 0,2,3,$4^2$,5,$6^2$,7,8,9,10,12 ~~~~~~
& ~~~~  0, 12 ~~~~ & ~~  0 ~~ & ~~ 0 ~~ & ~~ 0 ~~ \\ 
\hline
\end{tabular}\\[2pt]
\end{tiny}

\vspace{0.4in}

Table ~ 5 ~~ 
{{\em I} g.s. probabilities of four
fermions in a {\em j}=9/2 shell.}

\vspace{0.2in}

\begin{tiny}
\begin{tabular}{llllllllllll}
\hline
$I$           &  $0~~~$  &  $2~~~$  &  $3~~~$  &  $4~~~$   &
$5~~~$           & $6~~~$ & $7~~~$  &  $8~~~$  &  $9~~~$   & 
$10~~$          & $12~~$ \\
\hline
$  P^{exp}(I)$          &  66.4 &  3.7 &  0 &  11.8  &
0 ~                       &  0 ~~  &  0 ~  &  0.2~~~    &  0~~~  &
0  ~~~                       &  17.9 \\
$ P^{G_J=-1}(I)$          &  60 &   0 &   0 &  20~~~ &
0~~~~                       &  0~~~~  &  0~~~~  &  0~~~    &  0~~~  &
0~~~~                       &  20~~ \\
$ P^{G_J=\pm 1}(I)$          &  65.8~~ &  0.8 &  0.8 & 11.5~~ &
0.8~~                       &  1.5~~~  &  0.8~  &  0.8~~    &  0.8~~  &
0.8~                       &  15.8 \\  
\hline
\end{tabular}\\[2pt]
\end{tiny}

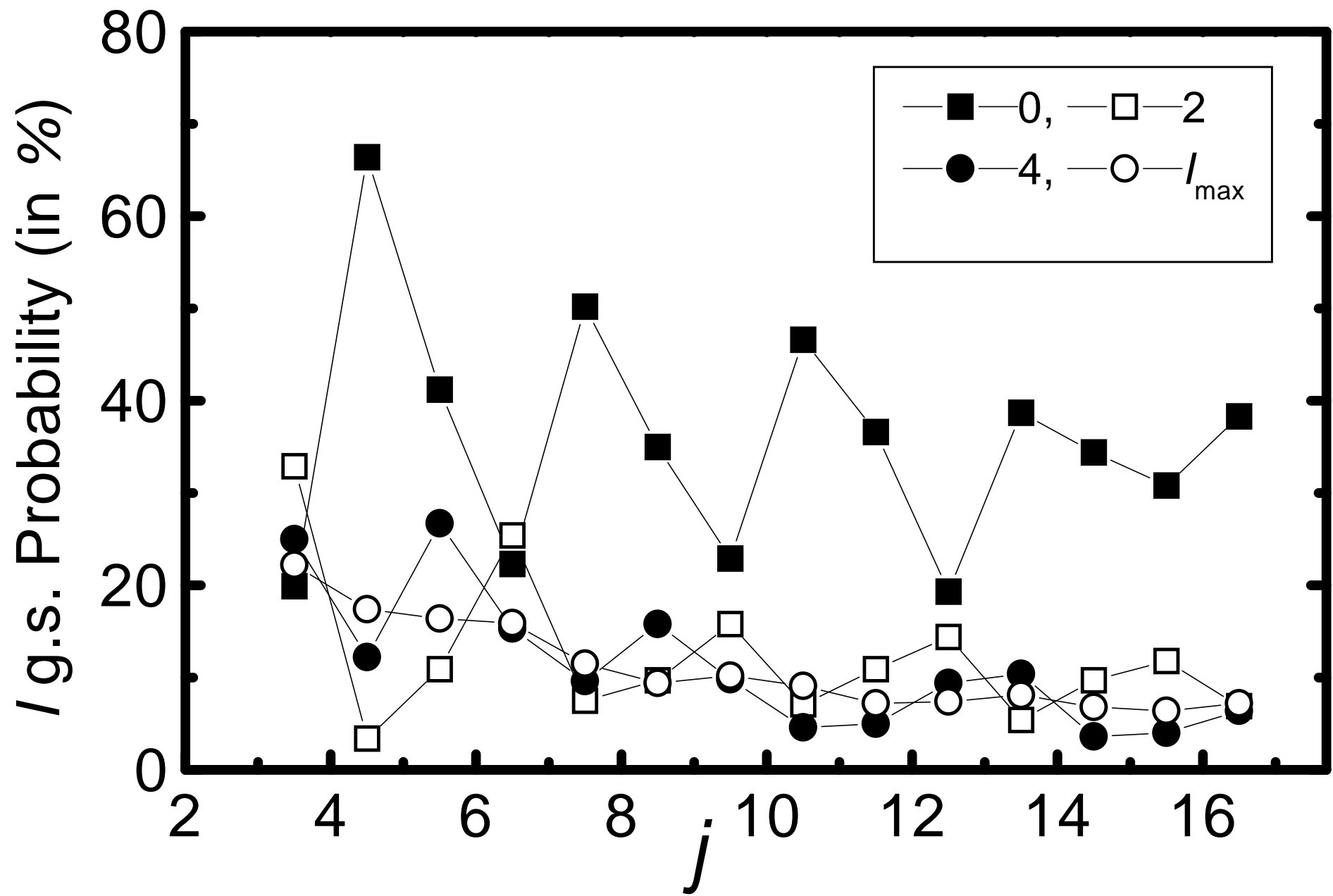
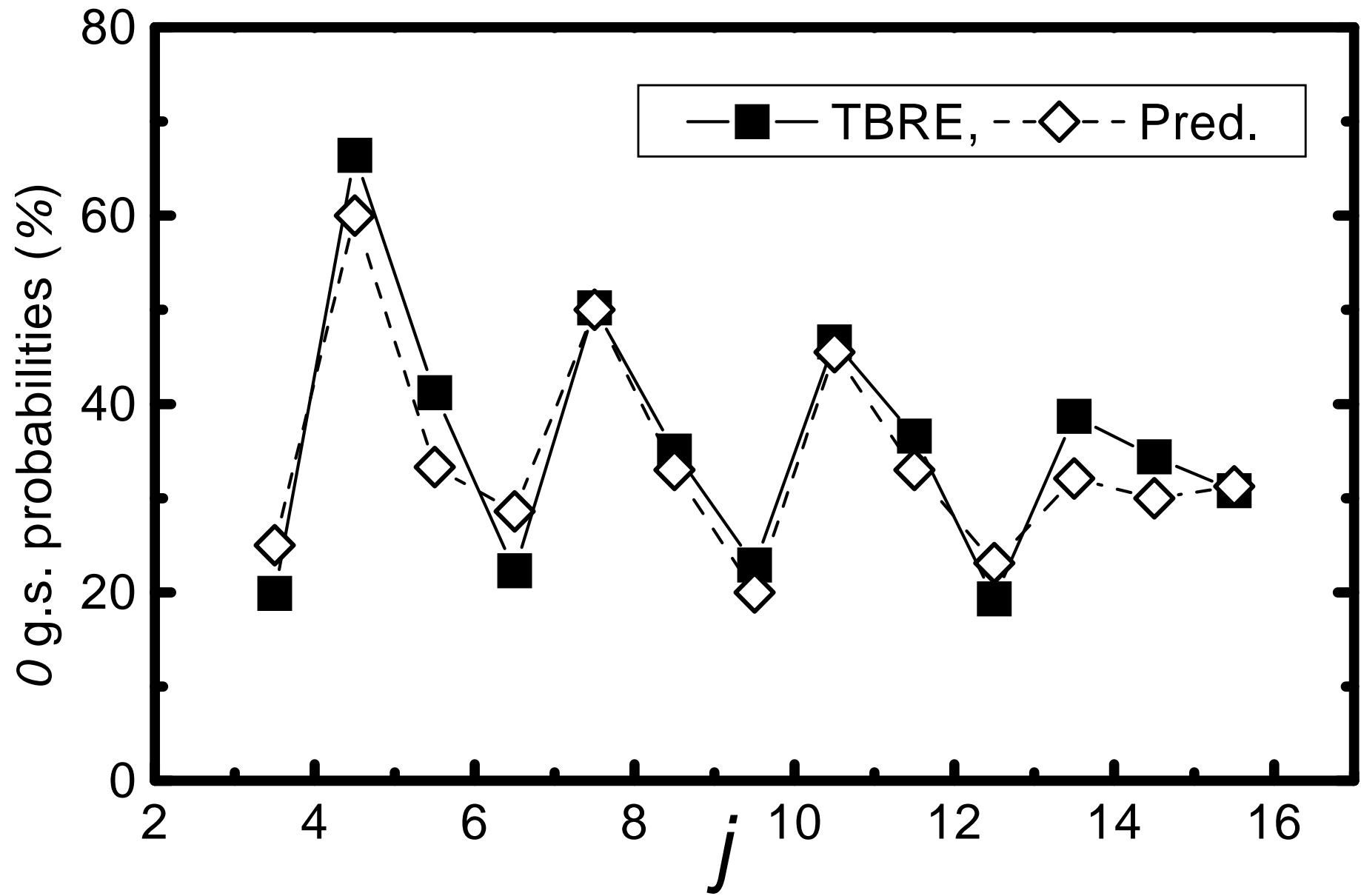
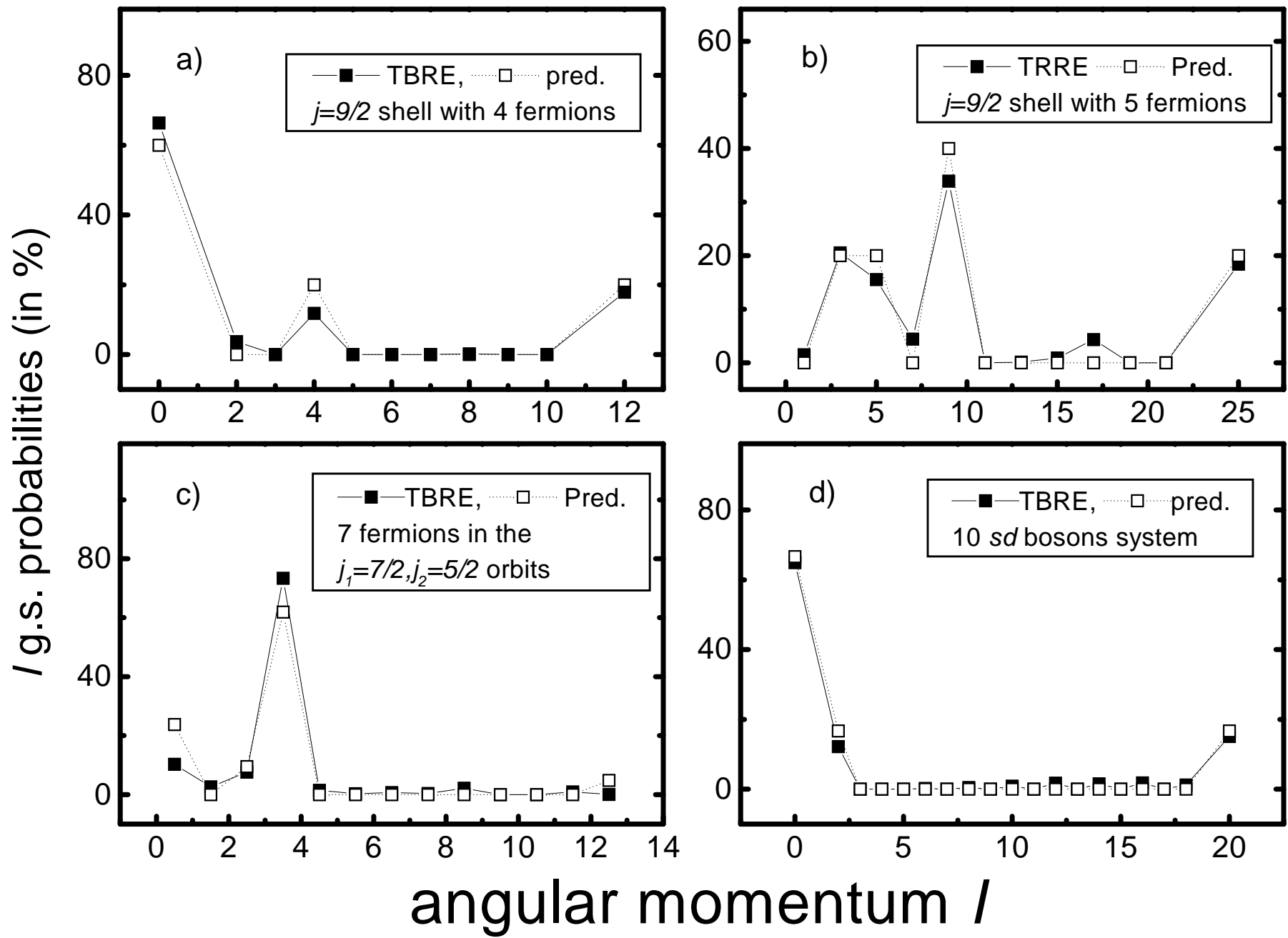
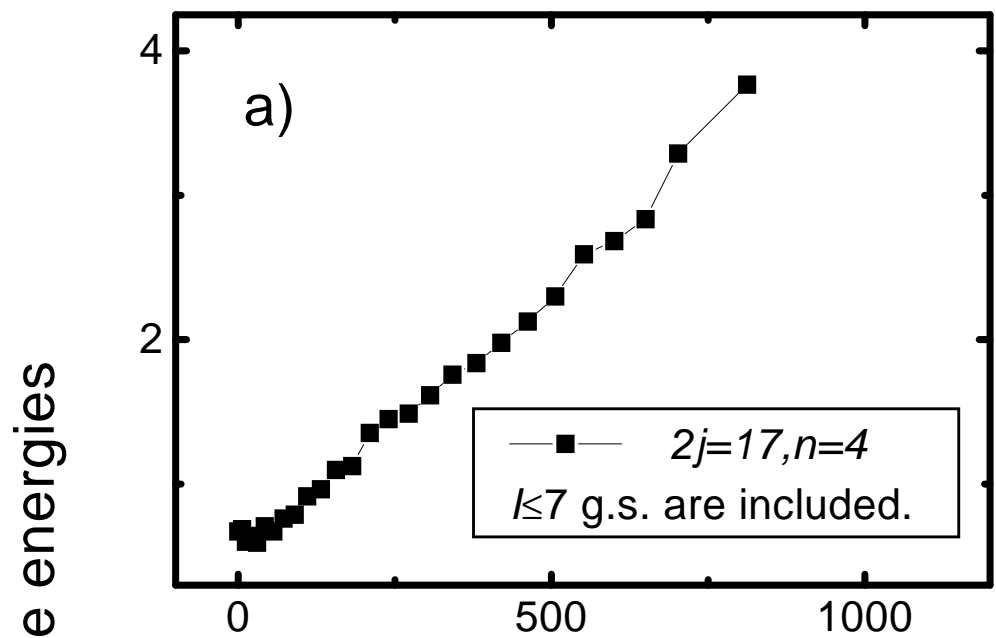
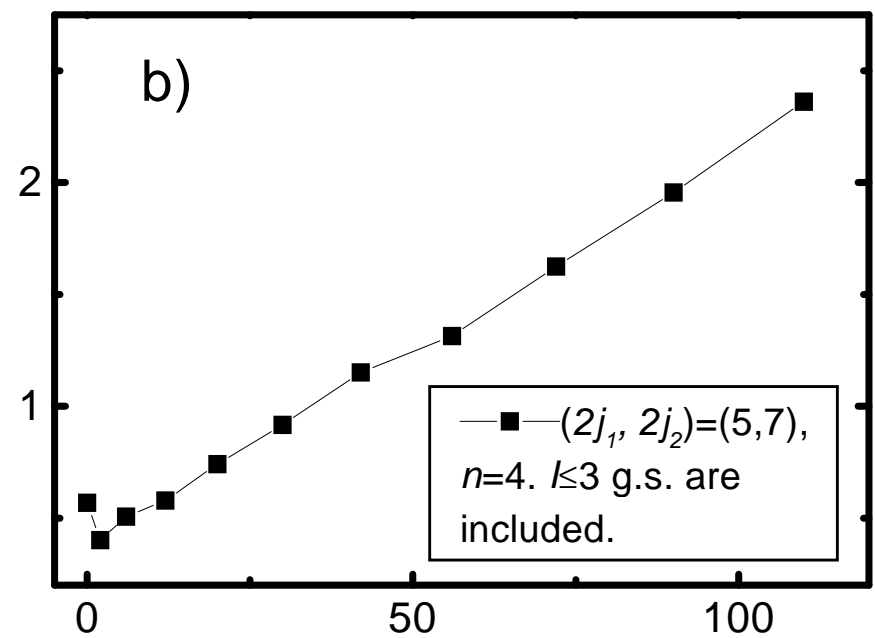
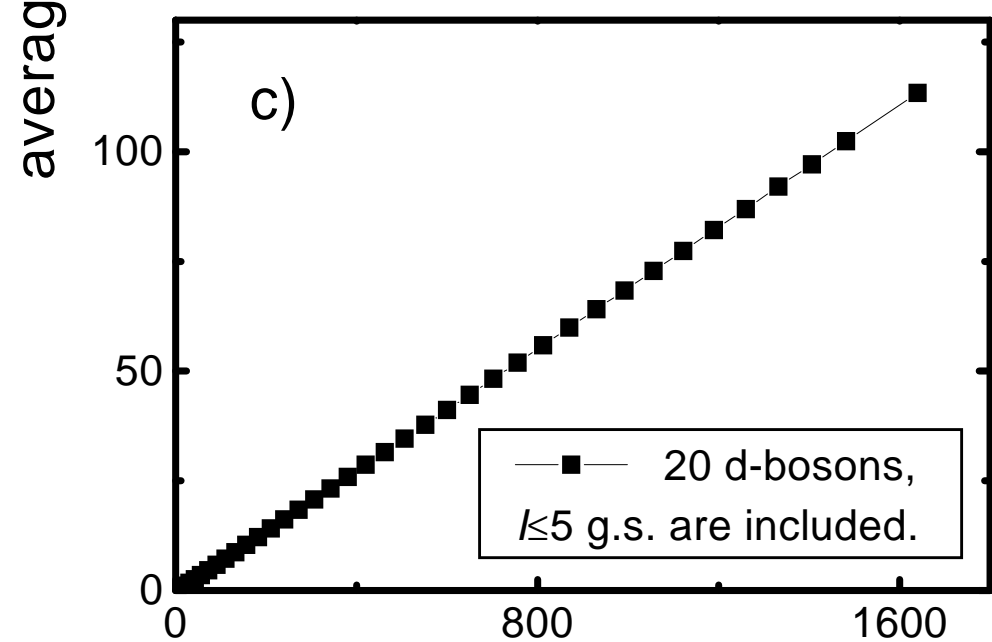
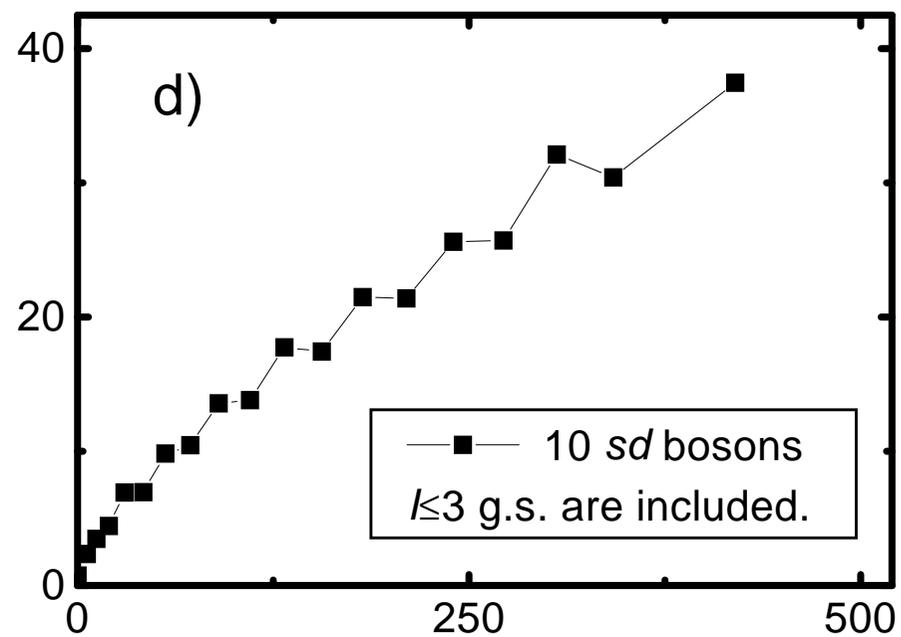
\end{document}